\documentstyle [12pt] {article}

\newcommand{\cs}[3]{{{#3} \brace {#1 #2}}}

\newcommand{\h}[1]{\mathop{\lambda}\limits_{#1}\ \!\!\!}

\newcommand{\edf}{\ {\mathop{=}\limits^{\rm def}}\ }

\newcommand{\al}{\alpha}

\begin{document}
\begin{center}
{ \bf{QUANTUM ROOTS IN GEOMETRY: II}}
\end{center}
\begin{center}
{\           }
\end{center}
\begin{center}{{\bf M. I. Wanas}\\
 Astronomy Department, Faculty of Science, \\ Cairo University,
 Giza, Egypt.\\ e-mail wanas@frcu.eun.eg}
\end{center}
\begin{abstract} The present work is a review of a series of papers,
published in the last ten years, comprising an attempt to find a
suitable avenue from geometry to quantum. It shows clearly that, any
non-symmetric geometry admits some built-in quantum features. These
features disappear completely once the geometry becomes symmetric
(torsion-less). It is shown that, torsion of space-time plays an
important role in both geometry and physics. It interacts with the
spin of the moving particle and with its charge. The first
interaction, {\bf{Spin-Torsion Interaction}}, has been used to
overcome the discrepancy in the results of the COW-experiment. The
second interaction, {\bf{Charge-Torsion Interaction}}, is similar to
the Aharonov-Bohm effect.

As a byproduct, a new version of Absolute Parallelism (AP) geometry,
the Parameterized Absolute Parallelism (PAP) geometry, has been
established and developed. This version can be used to construct
field theories that admit some quantum features. Riemannian geometry
and conventional AP-geometry are special cases of PAP-geometry.
\end{abstract}

\section{Introduction}
Space-time is not just a background in which physical interactions
take place. It is something more important than that. It interacts
with different properties of matter, while such properties affect
its geometric structure. In the context of Einstein's theory of
General Relativity (GR), it is widely agreed that gravity, the
weakest known fundamental interaction, is nothing but a space-time
property. This interaction cannot be switched off except in the case
of free falling frames. This is a direct consequence of the weak
equivalence principle (WEP). Due to the fact that gravity is the
weakest among the , so far, known interactions a majority of authors
ignore its effect in dealing with non-gravitational phenomena.

In view of  what is mentioned in the above paragraph, that gravity
is a space-time property, one cannot neglect space-time structure
whether gravity is weak or strong. In some cases, the neglect of
space-time structure leads to negative results or discrepancies
between theory and experiment. As a consequence, authors found
themselves obliged to add some peculiar assumptions in order to
remove such discrepancies or negative results. Such assumptions
might  result from a successful theory which takes space-time
structure into account. Several examples can be quoted here.  The
famous historical example, is the negative result of Michelson and
Morley experiment and the peculiar assumption of contraction of
lengths imposed, before special relativity, to interpret this
negative result.

The weak equivalence principle is verified experimentally, to an
accuracy of about $10^{-13}$ (cf.[1]). For this reason, many authors
deal with this principle as a law of nature rather than a suggestion
implying an empirical relation. In this context, let me quote
Eddington's statement [2] "{\it{...But the more important
consideration is the universality, rather than the accuracy, of the
experimental laws...}} ". The equations of motion, in the context of
GR, have been chosen in such a way that WEP is not violated. It is
well known that GR has been constructed in Riemannian geometry,
applying the philosophy of geometrization of physics. In the frame
work of this philosophy, Einstein has chosen the geodesic equation
to represent trajectories of test particles in gravitational fields.
This equation contains one geometric object characterizing the
geometry used, that is Christoffel symbol. It can be shown that,
since Christoffel symbol is a symmetric connection, it is always
possible to find a coordinate system in which any symmetric
connection vanishes at its origin. For this reason the WEP cannot be
violated in the context of GR written in Riemannian geometry.

On one hand, it is well known that the equation of motion in quantum
mechanics violates the WEP, as such equations contain terms
reflecting the intrinsic properties of the moving particle. On the
other hand, in the context of the geometrization philosophy motion
is described using paths of an appropriate geometry. Now, assume
that the quantum behavior, in the motion of elementary particle is a
result of some interaction between the quantum properties of such
particles and the background space-time.  Then, to look for quantum
properties in geometry, one has to examine path equations in
geometries other than the Reimannian one. In this framework, it is
obvious that:  \\
1. Riemannian geometry is not appropriate for describing the motion
of elementary particles, since its affine connection is symmetric.
\\
2. An appropriate geometry should admit a non-symmetric connection,
since the torsion part of connection is a tensor, that cannot be
removed using any coordinate transformation. Consequently, path
equations of such geometries, if used to describe trajectories of
test particles, would violate the WEP. If the geometrization
philosophy is a general philosophy that can be applied to the whole
physics, not only to gravity, then one would expect the appearance
of some quantum properties in the path equations of non-symmetric
type of geometries.

The paper is organized in the following manner. In section 2 we give
a brief account on the bases of  absolute parallelism (AP)-geometry,
and how to linearize their geometric objects. A brief summary of the
results displayed in the first part of this research is given in
section 3. In section 4 we give a review on recent results obtained
in the last few years. The paper is discussed and concluded in
section 5.
\section{Bases of AP-geometry}

 In the present section, we are going to give a brief summary of the
 bases of AP-geometry. This geometry is non-symmetric in the sense
that its affine connection is non-symmetric. As its path equations,
if used as trajectories of test particles, violate the WEP, one
would expect the appearance of some quantum properties in the motion
of such particles , if described in the context of this geometry.

AP-space is an n-dimensional manifold whose structure is defined
completely using n-linearly independent contravariant vector fields
 $\h{i}^{\mu}$, where the Latin indices  $(i=0,1,2,3,4,...,n) $ denote the vector number while
the Greek indices $( \mu = 0,1,2,3,4,.....n )$ denote the coordinate
component. The covariant components of such vector fields are
defined such that:
$$
 \h{i}^{\mu}\h{i}_{\nu} = \delta^{\mu}_{\nu} , \eqno{(1)}
$$
$$
 \h{i}^{\mu}\h{j}_{\mu} = \delta_{ij} , \eqno{(2)}
$$
summation convention is applied for both types of indices.  As a
direct consequence of the above mentioned linear independence, the
following linear connection can be defied,
$$
\Gamma^{\alpha}_{.\mu \nu} \edf \h{i}^{\mu}\h{i}_{\mu, \nu}.
\eqno{(3)}
$$
Using the building blocks of AP-space, one can define the following
symmetric tensors,
$$
g^{\mu \nu} \edf \h{i}^{\mu}\h{i}^{\nu} , \eqno{(4)}
$$

$$
g_{\alpha \beta} \edf \h{i}_{\alpha}\h{i}_{\beta} . \eqno{(5)}
$$
These tensors can be used as metric tensors, of a Riemannian
structure associated with the AP-space, when needed.

Now using the linear connection (3) we can define the torsion and
contortion tensors of the AP-space, respectively, as
$$
\Lambda^{\alpha}_{.\mu \nu} \edf \Gamma^{\alpha}_{.\mu \nu}-
\Gamma^{\alpha}_{.\nu \mu} \eqno{(6)}
$$

$$
\gamma^{\alpha}_{.\mu \nu} \edf \h{i}^{\alpha}\h{i}_{\mu ;\nu} .
\eqno{(7)}
$$
where (;) is used to characterize covariant differentiation using
Christoffel symbol $\cs{\mu}{\nu}{\alpha}$ defined, using (4)and
(5), in the usual manner. As a consequence of (3) and (7) we can
write
$$
\Gamma^{\alpha}_{.\mu \nu} \edf  \cs{\mu}{\nu}{\alpha} +
\gamma^{\alpha}_{.\mu \nu} .   \eqno{(8)}
$$
Contracting (6) and (7) we get the basic vector
$$
C_{\mu} \edf \Lambda^{\alpha}_{.\mu \alpha} = \gamma^{\alpha}_{.\mu
\alpha} . \eqno{(9)}
$$

Using the above mentioned tensors, Mikhail [3] has defined a set of
symmetric and antisymmetric second order tensors. These tensors are
found to be very useful in several physical applications. The
following table gives the definitions of these tensors.
\newpage
\begin{center}
 Table I: Second Order World Tensors [3]      \\
\vspace{0.5cm}
\begin{tabular}{|c|c|} \hline
 & \\
Skew-Symmetric Tensors                &  Symmetric Tensors   \\
 & \\ \hline
 & \\
${\xi}_{\mu \nu} \edf \gamma^{~ ~ \alpha}_{\mu \nu .
|{\stackrel{\alpha}{+}}} $ &
\\

${\zeta}_{\mu\nu} \edf C_{\alpha}~{\gamma^{~~ \alpha}_{\mu \nu .} }
$ &
\\
 & \\ \hline
 & \\
${\eta}_{\mu \nu} \edf C_{\alpha}~{\Lambda^{\alpha}_{.\mu \nu} } $ &
${\phi}_{\mu \nu} \edf C_{\alpha}~\Delta^{\alpha}_{.\mu \nu} $
\\

${\chi}_{\mu \nu} \edf \Lambda^{\alpha}_{. \mu
\nu|{\stackrel{\alpha}{+}} }$ & ${\psi}_{\mu \nu} \edf
\Delta^{\alpha}_{. \mu \nu|{\stackrel{\alpha}{+}}} $
\\

${\varepsilon}_{\mu \nu} \edf C_{\mu | {\stackrel{\nu}{+}}} - C_{\nu
| {\stackrel{\mu}{+}}}$ & ${\theta}_{\mu \nu} \edf C_{\mu |
{\stackrel{\nu}{+}}} + C_{\nu | {\stackrel{\mu}{+}}}  $
\\

${\kappa}_{\mu \nu} \edf \gamma^{\alpha}_{. \mu
\epsilon}\gamma^{\epsilon}_{. \alpha \nu} - \gamma^{\alpha}_{. \nu
\epsilon}\gamma^{\epsilon}_{. \alpha \mu}$   & ${\varpi}_{\mu \nu}
\edf  \gamma^{\alpha}_{. \mu \epsilon}\gamma^{\epsilon}_{. \alpha
\nu} + \gamma^{\alpha}_{. \nu \epsilon}\gamma^{\epsilon}_{. \alpha
\mu}$ \\
 & \\ \hline
 & \\
                  &  ${\omega}_{\mu \nu} \edf \gamma^{\epsilon}_{. \mu \alpha}\gamma^{\alpha}_{. \nu \epsilon}$   \\

                                      &  ${\sigma}_{\mu \nu} \edf \gamma^{\epsilon}_{. \alpha \mu} \gamma^{\alpha}_{. \epsilon \nu}$   \\

                                      &  ${\alpha}_{\mu \nu} \edf C_{\mu}C_{\nu}$   \\

                                      &  $R_{\mu \nu} \edf \frac{1}{2}(\psi_{\mu \nu} - \phi_{\mu \nu} - \theta_{\mu \nu}) + \omega_{\mu \nu}$          \\
 & \\ \hline
\end{tabular}
\end{center}
where $\Delta^{\alpha}_{. \mu \nu} \edf \gamma^{\alpha}_{. \mu \nu}
+ \gamma^{\alpha}_{.  \nu \mu} $ and $\Lambda^{\al}_{.~\nu \mu
|\stackrel{\sigma}{+}} \equiv
\Lambda^{\stackrel{\al}{+}}_{.~\stackrel{\mu}{+}{\stackrel{\nu}{+}}
| \sigma}$. It can be easily shown that there exist an identity
between skew-tensors, which can be written in the form [3],
$$\eta_{\mu\nu}+\varepsilon_{\mu\nu}-\chi_{\mu\nu}\equiv
0.$$ We see from Table I that the torsion tensor plays an important
role in the structure of AP-space in which all tensors in Table I
vanish when the torsion tensor vanishes.

The AP-geometry admits, at least, four linear connections: The
canonical connection (3), its dual $\tilde{\Gamma}^{\alpha}_{. \mu
\nu} (={\Gamma}^{\alpha}_{.\nu \mu} )$, its symmetric part
${\Gamma}^{\alpha}_{.(\mu \nu)}$ and the Christoffel symbol defined
 using (4) and (5). Using these connections, one can define the
following covariant derivative [4]
$$
A_{\stackrel{\mu}{+} }{}_{| \nu} \edf A_{\mu , \nu } -
A_{\alpha}\Gamma^{\alpha}_{. \mu \nu} , \eqno{(10)}
$$

$$
A_{\stackrel{\mu}{-}}{}_{| \nu} \edf A_{\mu , \nu } - A_{\alpha}
\tilde{\Gamma}^{\alpha}_{. \mu \nu} , \eqno{(11)}
$$

$$
A_{{\mu}| \nu} \edf A_{\mu , \nu } - A_{\alpha}\Gamma^{\alpha}_{. (
\mu \nu )} , \eqno{(12)}
$$

$$
A_{{\mu}; \nu} \edf A_{\mu , \nu } - A_{\alpha}
\cs{\mu}{\nu}{\alpha} . \eqno{(13)}
$$

The curvature tensors, corresponding to the above mentioned
connections, are thoroughly studied in a recent paper [5].

It has been shown [6] that the above 4-connections give rise to a
set of 3-path equations which can be written as:
$$
\frac{d^{2}x^{\mu}}{ds^{2-}} + \cs{\alpha}{\beta}{\mu}
\frac{dx^{\alpha}}{ds^{-}} \frac{dx^{\beta}}{ds^{-}} = 0,
\eqno{(14)}
$$

$$
\frac{d^{2}x^{\mu}}{ds^{2o}} + \cs{\alpha}{\beta}{\mu}
\frac{dx^{\alpha}}{ds^{o}} \frac{dx^{\beta}}{ds^{o}} = - \frac{1}{2}
\Lambda^{~ ~ ~ ~ \mu}_{(\alpha \beta) .}\frac{dx^{\alpha}}{ds^{o}}
\frac{dx^{\beta}}{ds^{o}} , \eqno{(15)}
$$
and
$$
\frac{d^{2}x^{\mu}}{ds^{2+}} + \cs{\alpha}{\beta}{\mu}
\frac{dx^{\alpha}}{ds^{+}} \frac{dx^{\beta}}{ds^{+}} = - \Lambda^{~
~ ~ ~ \mu}_{(\alpha \beta) .}\frac{dx^{\alpha}}{ds^{+}}
\frac{dx^{\beta}}{ds^{+}} . \eqno{(16)}$$  where $s^{-}, s^{o},
s^{+}$ are the affine parameters characterizing the corresponding
paths, respectively. These are the path equations admitted by the
AP-geometry. It is clear that particles following (15)or(16), as
physical trajectories, would violate the WEP. The R.H.S. of these
equations are tensors that cannot be removed using any coordinate
transformation. Further examination of these equations shows that
their R.H.S. represents a physical interaction between the torsion
of the background geometry and the quantum spin of the moving
particle. This will be displayed in some details in the next
section.

 In the context of the philosophy of geometrization  of physics, geometric
objects are used to represent physical fields. In some applications,
it is useful to expand such geometric objects using a certain
parameter, to study some physical phenomena. If we assume that
$$
\h{i}_{\mu} = \delta_{i \mu} + \epsilon h_{i \mu} \eqno{(17)}
$$
where $\delta$ is the Kroneckar delta, $h_{i \mu}$ are functions of
the coordinates and $\epsilon$ is a small parameter, then we can
expand the above mentioned objects using (17). The results of this
expansion are summarized in Table II [7]. In this table the symbol
$\bigotimes$ means that such terms are absent, while $\surd$
indicates the presence of these terms.
 \newpage
\begin{center}
 Table II: Orders of Magnitude of Geometric Objects of Physical
 Importance
\begin{tabular}{|c|c|c|c|c|} \hline
Object& zero order terms &$ 1^{st}$ order terms&   $2^{nd}$ order terms  & $3^{rd}$  and higher order terms\\
\hline
&&&& \\
$\h{i}_{\mu}$& $\surd $& $\surd$ & $\bigotimes$  & $\bigotimes$ \\
 $g_{\mu\nu}$&$ \surd$ &$ \surd$ &  $ \surd$ & $\bigotimes$ \\
$\h{i}^{\mu}$ & $ \surd$&$ \surd$ & $ \surd$  &$ \surd$ \\
$g^{\mu \nu}$ & $ \surd$&$ \surd$ & $ \surd$  &$ \surd$ \\
 \hline
 &&&& \\
  $\Gamma^{\alpha}_{\mu \nu}$& $ \bigotimes$&$ \surd$ & $ \surd$ &$
  \surd$\\
$\cs{\mu}{\nu}{\alpha}$ & $\bigotimes$&$ \surd$ & $ \surd$  &$
\surd$ \\
$\gamma^{\alpha}_{\mu \nu}$& $ \bigotimes$&$ \surd $& $ \surd $&
$\surd$\\ $\Lambda^{\alpha}_{\mu \nu}$& $ \bigotimes$&$ \surd$ & $
\surd $ &$ \surd$\\ $\Delta^{\alpha}_{\mu \nu}$& $ \bigotimes$&$
\surd$ & $ \surd$ &$
\surd$\\
$C_{\mu}$& $ \bigotimes$&$ \surd$ & $ \surd $&$ \surd$\\
 \hline
 &&&& \\
$\xi_{\mu \nu}$& $ \bigotimes$&$ \surd $& $ \surd $&$ \surd$\\
 &&&& \\
$\chi_{\mu \nu}$& $ \bigotimes$&$ \surd$ & $ \surd$ &$ \surd$\\
  &&&& \\
$\varepsilon_{\mu \nu}$& $ \bigotimes$&$ \surd $& $ \surd$ &$ \surd$\\
 &&&& \\
$\zeta_{\mu \nu}$& $ \bigotimes$&$ \bigotimes $& $ \surd $&$ \surd$\\
 &&&& \\
$\eta_{\mu \nu}$& $ \bigotimes$&$\bigotimes$ & $ \surd$ &$ \surd$\\
  &&&& \\
$\kappa_{\mu \nu}$& $ \bigotimes$&$ \bigotimes $& $ \surd$ &$ \surd$\\
\hline
&&&& \\
$\theta_{\mu \nu}$& $ \bigotimes$&$\surd$ & $ \surd$ &$ \surd$\\
  &&&& \\
$\psi_{\mu \nu}$& $ \bigotimes$&$\surd$ & $ \surd$ &$ \surd$\\
  &&&& \\
$\phi_{\mu \nu}$& $ \bigotimes$&$\bigotimes$ & $ \surd$ &$ \surd$\\
  &&&& \\
$\varpi_{\mu \nu}$& $ \bigotimes$&$\bigotimes$ & $ \surd$ &$ \surd$\\
  &&&& \\
$\omega_{\mu \nu}$& $ \bigotimes$&$\bigotimes$ & $ \surd$ &$ \surd$\\
  &&&& \\
  $\sigma_{\mu \nu}$& $ \bigotimes$&$\bigotimes$ & $ \surd$ &$ \surd$\\
  &&&& \\
$\alpha_{\mu \nu}$& $ \bigotimes$&$\bigotimes$ & $ \surd$ &$ \surd$\\
  &&&& \\
  \hline
  $R_{\mu \nu}$& $ \bigotimes$&$\surd$ & $ \surd$ &$ \surd$\\
  &&&& \\
$F_{\mu \nu}$& $ \bigotimes$&$\surd$ & $ \surd$ &$ \surd$\\
  \hline
\end{tabular}
\end{center}
where,
$$
{F}_{\mu \nu} \edf {C}_{\mu, \nu} - {C}_{\nu, \mu}.$$

\section{A Brief Review  of Part I}
  In the previous section, it is shown that AP-geometry admits three
  path equations while the affine connections are, at least four.
It can be shown [8] that, using  connections different from the
above mentioned and defining the corresponding path equations we
always get one or more of the equations belonging to the above set.
We consider this property as an important feature of non-symmetric
geometries. Another important feature appeared in the set of
equations (14),(15)and (16), that is the coefficient of the torsion
terms jumps, with step one-half from an equations to the next. We
consider this last feature as reflecting some quantum properties
 of the paths admitted by any non-symmetric geometry. This
has been tempting to generalize the AP-geometry in order to be used
in describing some quantum physical phenomena. The new version is
known as the Parameterized Absolute Parallelism (PAP)-geometry [9],
which is characterized by the linear connection,
$$
\nabla^{\alpha}_{. \mu \nu} \edf \cs{\mu}{\nu}{\alpha} + b
\gamma^{\alpha}_{. \mu \nu}, \eqno{(18)}
$$
where $b$ is a dimensionless parameter. The parameterized path
equation, corresponding to the linear connection (18), can be
written as

$$
\frac{d^{2}x^{\mu}}{d\tau^{2}} + \cs{\alpha}{\beta}{\mu}
\frac{dx^{\alpha}}{d\tau} \frac{dx^{\beta}}{d\tau} = -b \Lambda^{~ ~
~ ~ \mu}_{(\alpha \beta) .}\frac{dx^{\alpha}}{d\tau}
\frac{dx^{\beta}}{d \tau} . \eqno{(19)}
$$
where $Z^{\mu} (= \frac{dx^{\mu}}{d\tau})$ is the  vector tangent to
the path and $\tau$ is the parameter varying along the path. The
R.H.S. of equation (19) is suggested to represent a type of
interaction between the quantum spin of the moving particle and the
torsion of the background space-time. For  some physical and
dimensional considerations, the parameter $b$ is suggested to take
the form [10]
$$
 b = \frac{n }{2}\alpha \gamma   \eqno{(20)}
$$
where $n$ is a natural number taking the values $0,1,2,3,...$ for
particles with spins $0, \frac{1}{2}, 1,...$ respectively, $\alpha$
is the fine structure constant and $\gamma$ is a dimensionless
parameter to be fixed by experiment or observation.

Using equation (17) to expand the quantities in (19) and considering
the results given in Table II, it has been shown that the linearized
version of (19) gives rise to [10]

$$
\Phi_{s} = \Phi_{N}(1- \frac{n }{2}\alpha \gamma) \eqno{(21)}
$$
where $\Phi_{N}$ is the Newtonian gravitational potential and
$\Phi_{s}$ is the gravitational potential felt by a spinning
particle. Slow motion assumption is imposed on the tangent vector
$Z^{\mu}$.

Using the reduced gravitational potential $\Phi_{s}$ in the
calculations of the trajectories of thermal neutrons (spin
$\frac{1}{2}-$particles), it is shown that the discrepancy in the
COW-experiment can be removed [11]. Also, a satisfactory
interpretation for the time delay of massless spinning particles
(neutrinos, photons and gravitons !) detected from SN1987A, has been
given [12].

This is a brief summary of the first part of the work [13].
\section{Review of Some Recent Developments}
As mentioned in the section 3, there are some experimental and
observational evidence for the existence of spin-torsion
interaction. On the other hand, equation (19) gives the effect of
this interaction on the trajectory of a spinning particle. It is of
importance to connect this interaction to some observable
quantities. For this reason, the effect of the spin-torsion
interaction, on the value of Chandrasekhar limit, is obtained [14].
The formula giving this value can be written as
$$
M_{s} = M_{ch}(1 + \frac{3}{4} \alpha  \gamma ) \eqno{(22)},
$$
where $M_{s}$  and $M_{ch}$ are the spin-dependent and spin
independent Chandrasekher masses respectively. Formula (22) is
derived using the reduced gravitational potential (21), so it can be
applied to white dwarfs ( and can not be applied to strong fields
e.g. the field of neutron stars). An observable quantity connected
to (22) is the gravitational red-shift, which can be written as
$$
Z_{s} = Z_{ch}(1 + \frac{1}{2} \alpha \gamma)     \eqno{(23)}. $$
The results obtained are listed in the following Table III for the
star 40 Eri B.

\begin{center}
 Table III: Red-Shift for 40 Eri B [14]
  \end{center}
 \begin{center}
\begin{tabular}{|c|c|c|c|} \hline
&&& \\
parameter& observed $(Z)$ & Spin-independent $(Z_{ch})$ & Spin-dependent $(Z_{s})$     \\
&&& \\
\hline
& & &  \\
 Red-shift $Rms/s$ & $23.9 \pm 1.3$ & $22.0\pm 1.4$ & $22.1
\pm 1.4$  \\
& & &  \\
\hline
\end{tabular}
\end{center}
From Table III it is clear that the spin dependent value, calculated
using (22), is more close to the observed value than the spin
independent value. For further details see ref. [14] and the
references listed therein.

It has been shown [15] that, although the quantum roots appear
firstly in the path equations (through the affine connection) of
PAP-geometry, these roots can diffuse and distribute among other
geometrical entities, e.g. torsion, curvature..... Furthermore, the
PAP-geometry admits a number of PAP-spaces, each of which is
characterized by a certain connection, torsion and curvature. The
Riemannian case and the conventional AP-case are just special cases
of PAP-geometry corresponding to $b=0, b=1$, respectively. A
particle having a certain quantum spin will feel certain connection,
torsion and curvature. Consequently, its trajectory will belong to
one of the PAP-spaces, depending on its spin. This shows how some
quantum roots are built in non-symmetric geometries.

It is well known that, for each path equation in a certain space,
there exits a corresponding path deviation equation. Such deviation
equations are of certain importance in applications. The path
deviation equations, corresponding to the set of path equations in
the AP-geometry (14), (15) and (16), are derived in [16]. These path
deviation equations can be represented by:
$$
\frac{D^{2}\Psi^{\alpha}}{D {\tau}^{2}} + a ( R^{..\alpha .}_{\mu
\beta .\rho} + \tilde{Q}^{..\alpha .}_{\mu \beta .\rho}) U^{\mu}
\Psi^{\beta} U^{\rho} = b \Lambda^{.. \alpha}_{\nu \rho .} \frac{D
\Psi^{\nu}}{D \tau} U^{\rho} + c \gamma^{\alpha}_{. \rho \nu}
\frac{D \Psi^{\nu}}{D \tau}U^{\rho}, \eqno{(24)} $$ where
$R^{\alpha}_{. \beta \gamma \delta}$ is the Riemann-Christoffel
curvature tensor and $\Psi^{\alpha}$ is the deviation vector,
$U^{\mu} \edf \frac{d x^{\mu}}{d \tau}$, $\tau$ is the parameter
characterizing the path,
$$
\tilde{Q}^{\lambda}_{. \mu \rho \sigma} \edf \gamma^{\lambda}_{ \rho
\mu;\sigma} - \gamma^{\lambda}_{ \sigma \mu;\rho} +
\gamma^{\epsilon}_{\rho \mu} \gamma^{\lambda}_{. \sigma \epsilon} -
\gamma^{\epsilon}_{. \sigma \mu} \gamma^{\lambda}_{. \rho \epsilon}
\eqno{(25)}
$$
and $a, b, c$ are parameters whose values are given in Table IV

\begin{center}
{Table IV: Values of the Parameters of Equation (24) } \\
\end{center}
\begin{center}
\begin{tabular}{|c|c|c|c|} \hline
 ~~~~Connection used~~~~  &~~~~ a~~~~ &~~~~ b~~~~ &~~~~ c~~~~  \\
 \hline
& & &\\
 ${\Gamma}^{\mu}_{.~\alpha \nu} $ &~0~&~1~&~0    \\
\hline
& & &\\
  ${\Gamma}^{\mu}_{.~ (\alpha \nu )}$ &~$\frac{1}{2}~$ &~  $\frac{1}{2}$ ~& ~1      \\
   \hline
& & & \\
  $\tilde{\Gamma}^{\mu}_{.~\alpha\nu } $ &~1~&~0~&~2      \\
   \hline
\end{tabular}
\end{center}
From this table it is clear that the jumping values appear, not only
in the torsion term ( parameter $b$) but also in the curvature
(parameter $a$ ) and in the contortion (parameter $c$) terms. Table
IV shows also that while the parameters $ (a) \& (b) $ jump with
step of one-half (half integer), the jumping step of the parameter
$c$ is one (integer).

 As stated above, the path equations (19) with
quantum features, has been used to describe trajectories of spinning
particles in gravitational fields. The term on the R.H.S. of this
equation is suggested to represent a type of interaction between
spin and torsion. Now, what about the motion of a charged particle
in a background of gravity and electromagnetism? It has been shown
[17], in the context of the generalized field theory (GFT) [18], a
theory unifying gravity and electromagnetism, that the basic vector
(9) plays the role of Maxwell's electromagnetic potential. Modifying
the Lagrangian function used to obtain the set (15), (16) and (17),
by introducing the vector (9), we get [19] a set of three
 new path equations, which can be written in the general form:
$$
{\frac{dZ^\mu}{d\tau}} + a_{1}~ \cs{\alpha}{\beta}{\mu} Z^\alpha
Z^\beta = -~a_{2}~ \Lambda^{~ ~ ~ ~ \mu}_{(\alpha \beta) .}
~~Z^\alpha Z^\beta -~a_{3}~ \hat{F}^{\mu}_{. \nu}Z^{\nu} -~a_{4}~
g^{\mu \delta}\hat{C}_{_{\stackrel{\nu}{+}} | \delta}Z^{\nu} -~
a_{5}~ \Lambda^{~ ~ ~ ~ \mu}_{\alpha \beta .} ~~\hat{C}^{\alpha}
Z^\beta , \eqno{(26)}
$$
where $Z^{\mu}$ is the tangent of the path (26),
$$
\hat{F}_{\mu \nu} \edf \hat{C}_{\mu, \nu} -\hat{C}_{\nu, \mu} = l
{F}_{\mu \nu},$$
$$
\hat{C}_{\mu} \edf l C_{\mu},
$$
 $l$ is a dimensional parameter and  $a_{1}, a_{2}, a_{3}, a_{4},
$ and $a_{5}$ are numerical parameters whose values are listed in
 Table V.
\newpage
\begin{center}
 {Table V: Values of the Parameters of Equation (26)}
\end{center}
\begin{center}
\begin{tabular}{|c|c|c|c|c|c|} \hline
 Connection used&$a_{1}$&$a_{2}$& $a_{3}$& $a_{4}$ & $a_{5}$  \\
 \hline
& & & & &  \\
  ${\Gamma}^{\alpha}_{.~\mu \nu} $ & 1& 1& 1& 1 & 1       \\
& & & & &  \\
\hline
& & & & &  \\
   ${\Gamma}^{\alpha}_{.~ (\mu \nu )}$ &1 &  $\frac{1}{2}$ & 1 & 1 & $\frac{1}{2}$     \\
& & & & &  \\
   \hline
& & & & &  \\
  $\tilde{\Gamma}^{\alpha}_{.~\mu \nu } $ & 1&  0 & 1 & 1& 0   \\
& & & & &  \\
 \hline
\end{tabular}
\end{center}
 From Table V it is clear that two parameters $(a_{2}, a_{5} )$ have jumping values,
  while the other three parameters have the same  (non- jumping) value. Using the terminology
 of the present work, one can consider $(a_{2}, a_{5} )$ as quantized parameters. The term whose coefficient is
  $a_{3}$ represents the well known Lorentz force. Taking $l=0$ ( $l$ includes the electric
  charge of the moving particle), one can switch off the last three terms of the equation (26).
   This will reduce the set (26) to the old set (14), (15) and (16).

A striking result is that the last term in (26) represents a direct
influence of the electromagnetic potential on the trajectory of a
charged particle.  Note that, this term is quantized and gives rise
to a phenomenon similar to the Aharonov-Bohm effect which cannot be
obtained through any known classical treatment. Also, note that this
term will not vanish even if the Lorentz force vanishes (as a result
of vanishing of the electromagnetic field tensor $F_{\mu \nu}$). The
last term of (26) can be considered to represent a type of
interaction between the charge of the moving particle and the
torsion of the background space-time. We call this interaction the
"{\bf{Charge-Torsion Interaction}}". This represents another type of
interactions between an intrinsic property of the moving particle
and the background space-time structure.
\section{Discussion and Conclusion}
In this paper, it is shown that non-symmetric geometries are capable
of interpreting some quantum behaviors of elementary particles in a
background of gravity and electromagnetism. We are not doing quantum
mechanics in its conventional form, but we have studied the effect
of the quantum properties of an elementary particle on its
trajectory, using geometry. This paper throws some light on the
possibility of dealing with some quantum phenomena starting from
pure geometric considerations.

Using an appropriate geometry, i.e. a geometry with simultaneously
non-vanishing curvature and torsion, it is shown that some
interactions, between the torsion and the intrinsic quantum
properties of the moving particles, can be discovered and
interpreted. The interactions discovered, so far, using this scheme
are the {\bf{Spin-Torsion Interaction}} and {\bf{Charge-Torsion
Interaction}}. The later interaction gives rise to a phenomenon
similar, if not identical, to the Aharonov-Bohm effect, which can
not be accounted for using any known classical treatment. Equation
(26) gives rise to both interactions in addition to the Lorentz
force. It appears clearly from this equation, that Lorentz force is
not quantized while the charge-torsion interaction (the
Aharonov-Bohm effect) is quantized. This result is in agreement with
experience in physics and with experiment.

The results obtained, so far, in this new direction concerning the
relation between quantum and geometry, illuminate the importance of
taking space-time structure in dealing with non-gradational
phenomena. Furthermore, one can draw the following conclusion:
"{\it{Physical phenomena,  considered so far, in the application of
the present new trend, can be interpreted as interaction between
space-time properties (especially torsion) and the intrinsic
properties of elementary particles}}". Such type of interaction may
be used to interpret the accelerating expansion of our Universe and
may give a possible source for "{\bf{Dark Energy}}".

The author would like to thank the SOC for inviting him to give this
talk.
\section*{References}
{[1]} C.M. Will (2006) Living Rev. Relativity {\bf{9}}, 3.  \\
{[2]} A.S. Eddington (1923) {\it{The Mathematical Theory of
Relativity}}, Cambridge Univ. Press. \\
{[3]} F.I. Mikhail (1952) Ph.D. Thesis, London University. \\
{[4]} M.I. Wanas (2001) Stud. Cercet. \c Stiin\c t. Ser. Mat. Univ.
Bac\u au {\bf{10}}, 297;

 gr-qc/0209050 \\
{[5]} N.L. Youssef and A. Sid-Ahmed (2006) gr-qc/0604111.  \\
 {[6]} M.I. Wanas, M. Melek and M.E. Kahil (1995) Astropys. Space Sci. {\bf{228}} ,273;

 gr-qc/0207113. \\
{[7]} M.I. Wanas (1975) Ph.D. Thesis, Cairo University. \\
 {[8]} M.I. Wanas and M.E. Kahil (1999) Gen. Rel. Grav.,
{\bf{31}}, 1921;  gr-qc/9912007. \\
 {[9]} M.I. Wanas (2000) Turk. J. Phys., {\bf{24}}, 473; gr-qc/0010099. \\
{[10]} M.I. Wanas (1998) Astrophys. Space Sci.,
{\bf{258}}, 237; gr-qc/9904019. \\
  {[11]}M.I. Wanas, M. Melek, and M. E. Kahil
(2000) Gravit. Cosmol.,
{\bf{6} }, 319;

gr-qc/9812085. \\
{[12]} M.I. Wanas, M. Melek, and M.E. Kahil (2002) Proc. MG IX, part
B, p.1100,

Eds.
V.G. Gurzadyan et al. (World Scientific Pub.); gr-qc/0306086. \\
{[13]} M.I. Wanas (2002) Proc. XXV International Workshop on "{\it
Fundamental Problems

of High Energy Physics and Field Theory}", held
in Protvino, Russia, 24-28 June 2002,

ed. V.Petrov; gr-qc/0506081.\\
{[14]} M.I. Wanas (2003) Gravit. Cosmol. {\bf{9}}, 109. \\
 {[15]}  M.I. Wanas (2003) Algebras, Groups and Geometries,
{\bf20}, 345a.   \\
{[16]} M.I. Wanas and M.E. Kahil (2005) Proc. MG X, part B, p.1440,
Eds.

M. Novello et al. (World Scientific Pub.); gr-qc/0605036. \\
 {[17]} F.I. Mikhail, and M.I. Wanas (1981) Int. J. Theoret.
Phys. {\bf{20}}, 671. \\
{[18]} F.I. Mikhail and M.I. Wanas (1977), Proc. Roy. Soc.
Lond. {\bf{A 356}}, 471. \\
{[19]} M.I. Wanas and M.E. Kahil (2005) Int. J. Geom. Methods Mod.
Phys.
{\bf{2}}, 1017;

gr-qc/0408029. \\
\end{document}